\documentclass[]{icas2024} 
\def\BibTeX{{\rm B\kern-.05em{\sc i\kern-.025em b}\kern-.08em
    T\kern-.1667em\lower.7ex\hbox{E}\kern-.125emX}}

\usepackage{graphicx}
\usepackage{caption}
\usepackage{subcaption}
\usepackage{amsmath}
\usepackage[version=4]{mhchem}
\usepackage{siunitx}
\usepackage{longtable,tabularx}
\usepackage{multirow}
\usepackage[nolist,nohyperlinks]{acronym}
\usepackage{todonotes}
\usepackage{svg}
\TitlePaper{EXPLORING MULTI-FIDELITY AEROELASTIC TAILORING: \\ PROSPECT AND MODEL ASSESSMENT}
\AuthorPaper[1]{Hauke Maathuis}
\AuthorPaper[2]{Saullo G.P. Castro}
\AuthorPaper[3]{Roeland De Breuker}

\affil[1]{Ph.D. Candidate, Faculty of Aerospace Engineering, Delft University of Technology, h.f.maathuis@tudelft.nl}
\affil[2]{Associate Professor, Faculty of Aerospace Engineering, Delft University of Technology, s.g.p.castro@tudelft.nl}
\affil[3]{Associate Professor, Faculty of Aerospace Engineering, Delft University of Technology, r.debreuker@tudelft.nl}

\abstractEnglish{The design and optimisation of aircraft wings are critical tasks in aerospace engineering, requiring a balance between structural integrity, aerostructural performance, and manufacturability. This multifaceted challenge involves the interplay of various disciplines, each with distinct parameters and constraints. Traditional design approaches often fall short, necessitating advanced methodologies like Multidisciplinary Design Optimisation (MDO). MDO integrates aerodynamic, structural, and manufacturability analyses to explore a vast design space and identify optimal solutions that meet performance, safety, and cost criteria.\\
Advancements in manufacturing technologies and material sciences have led to the increased use of composite materials, which offer an excellent weight-to-strength ratio. Aeroelastic Tailoring, which incorporates directional stiffness into structural design, further enhances performance. This study employs lamination parameters to efficiently represent composite layups within a gradient-based optimisation process, aiming to minimise weight while ensuring feasibility across multiple constraints.\\
The work highlights the challenge of optimising aircraft designs using multiple models of varying fidelity. Traditional sequential optimisation approaches, which progressively integrate disciplines, may miss potential superior designs due to limited initial information. Instead, concurrent optimisation schemes are explored, utilising both low-fidelity (beam-based) and high-fidelity (shell-based) models. This approach promises structural feasibility, reduces computational costs, and incorporates high-fidelity information early in the design process.\\
The envisioned methodology bridges different design stages, enabling better overall aircraft performance. By aligning and comparing a beam-based and shell-based model, the study explores their use in multi-fidelity optimisation. The results demonstrate the feasibility and benefits of this approach, offering a robust framework for future aircraft design projects.}

\begin{document}

\body  

\section{Introduction}\label{ch:intro}

\begin{figure}[ht]
	\centering
	\includegraphics[width=9cm]{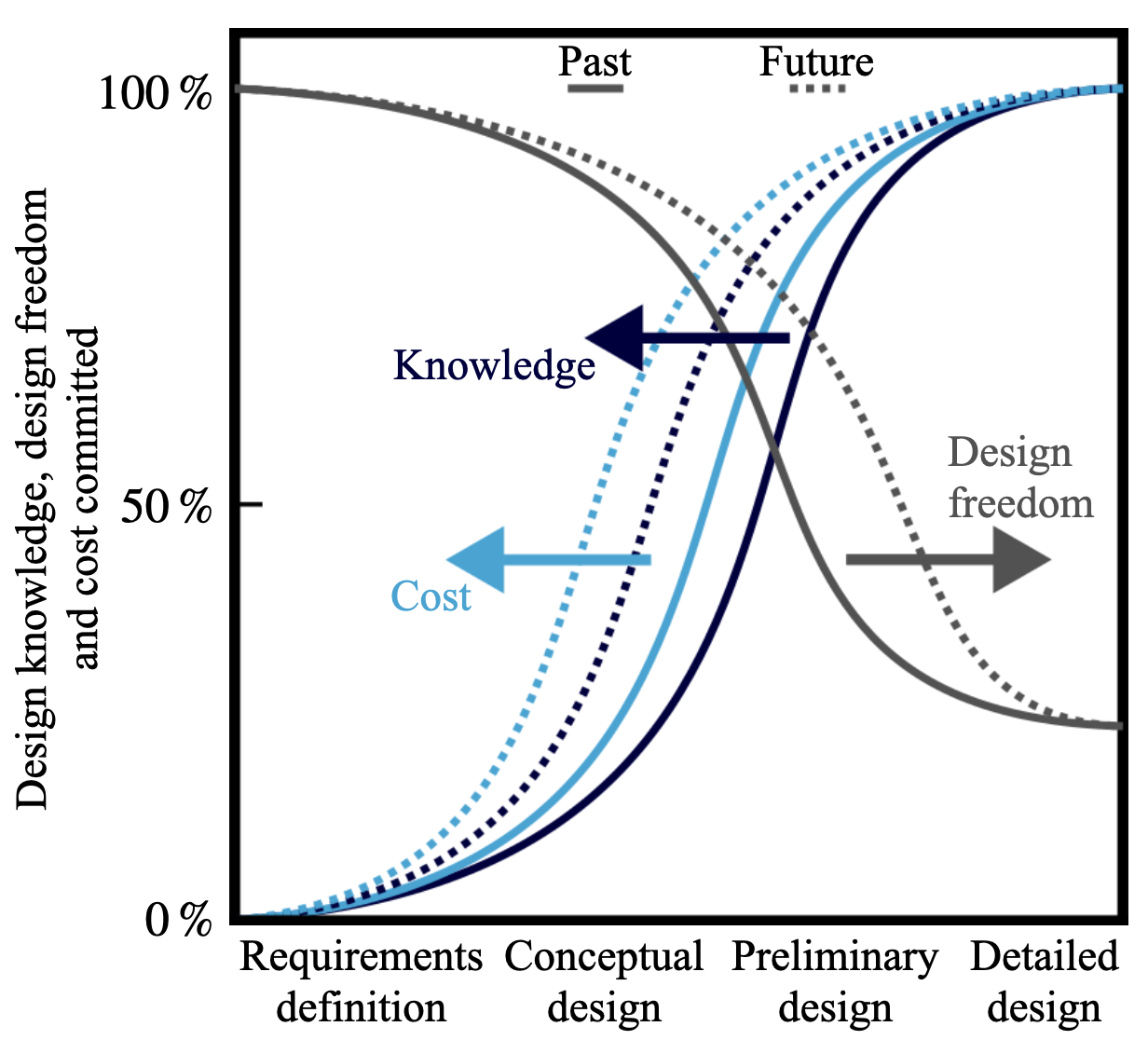}
	\caption{Life-Cycle Design Stages (adopted from \cite{1998Mavris}).}
	\label{fig:lifecycle}
\end{figure}

The design and optimisation of aircraft wings stand at the forefront of aerospace engineering, demanding a meticulous balance between structural integrity, aerostructural performance, and manufacturability. The complexity of this task is amplified by the interplay of multiple disciplines, each governed by its own set of parameters and constraints. Traditional design approaches often fall short in addressing this multi-faceted challenge, necessitating the adoption of advanced methodologies that can seamlessly integrate diverse disciplinary insights.\\
The increase in performance, or in other words a weight reduction of the structure, leads to lighter and slender wings which pose challenges like low natural frequencies coupling with aircraft motions or structural failures like buckling, due to thin-walled design. Early consideration of these disciplines in aircraft development using Multidisciplinary Design Optimisation (MDO) ensures an optimal yet safe structure. MDO emerges as a powerful framework to tackle this complexity, offering a holistic approach that simultaneously considers the myriad factors influencing wing design. By integrating aerodynamic, structural, and manufacturability analyses, MDO enables the exploration of a vast design space to identify solutions that meet performance, safety, and cost criteria.\\
Thanks to the advancements in manufacturing technologies and material sciences, composite materials are being used increasingly more due to their excellent weight-to-strength ratio. A promising technology to improve the performance of aircraft by optimising the structure is the application of Aeroelastic Tailoring. Thereby, directional stiffness is embodied into the structural design to control aeroelastic deformation to beneficially impact on the aerodynamic and structural performance of the aircraft \cite{1986Shirk}. Within this work, the concept of lamination parameters provides a robust means of representing composite layups. Lamination parameters abstract the detailed layup configuration into a set of variables that can be efficiently handled within a (gradient-based) optimisation process. By representing the stiffness and strength characteristics of composite materials in a concise form, lamination parameters enable the precise tailoring of material properties to meet specific design requirements while lowering the number of design variables to be considered by the optimiser. However, the use of lamination parameters demands a subsequent discrete optimisation to retrieve stacking sequences  \cite{2013Dillinger}. \\
Thus, in the following work, the emerging mathematical problem can be summarised as finding a combination of lamination parameter and thicknesses of the panels to obtain the minimum weight while ensuring feasibility, evaluated based on $G$ multidisciplinary constraints, formulated as 
\begin{equation}\label{eq:generaloptimisationproblem}
	\min_{\textbf{x} \in \mathcal{X}} f(\textbf{x}) \ \text{s.t.} \forall i \in \{1,...,G\}, c_i(\textbf{x}) \leq 0
\end{equation}
where $f(\textbf{x})$ represents the weight as the objective function, $\textbf{x} = \{ \textbf{x}_{lp}, \textbf{x}_{t} \} \in \mathcal{X} \subset \mathbb{R}^D$ is the $D$-dimensional vector of design variables within the feasible design space, consisting of $\textbf{x}_{lp}$ as the design variables representing lamination parameters and $\textbf{x}_{t}$ the thicknesses. \\
Typically in aircraft design, multiple models exist, each relying on different formulations, such as beam or shell formulations for structural analysis and panel codes or CFD analysis for aerodynamics. The process of aircraft development usually begins with a pre-conceptual design from which an initial configuration is derived. Following this, preliminary sizing is conducted using a simplified model (hereafter referred to as the low-fidelity model). Even at this early stage, it is desirable to involve as many disciplines as possible to achieve an optimal design. However, this is often computationally infeasible. Instead, current design processes employ a sequential approach where disciplines are gradually integrated or discarded as needed \cite{1998Mavris}.
This situation presents a significant challenge. Suppose a gradient-based optimisation identifies the local optimum in the initial design phase, taking into account disciplines that are too costly to consider at a later stage. When these outcomes are applied to a more intricate model, aiming to optimise additional design features and incorporating a different set of disciplines, potential superior designs for enhancing aircraft performance might be overlooked due to limited information and the "freezing" of the previous design optimisation. The sequential optimisation approach stems from the need for detailed models. Attempting to optimise all design variables and disciplines simultaneously (such as structural details, flight dynamics, etc.) is computationally impractical \cite{Kennedy2014, Kennedy2014-2}. Hence, multiple models are employed across distinct design phases.\\
Despite the utilisation of lamination parameters to condense the number of design variables, the Aeroelastic Tailoring problem remains high-dimensional, with a common design problem easily involving hundreds or thousands of design variables. These disciplines are accounted for by formulating $G$ constraints for the optimisation problem (see Equation \ref{eq:generaloptimisationproblem}), arising from the multitude of analyses mentioned earlier. In such cases, the number of constraints is very large as well (when not relying on aggregation methods), further complicating the optimisation. This is why current design processes apply gradient-based local optimisation methods.\\
The present work addresses this problem by aiming to use multiple models of differing fidelity in a concurrent optimisation scheme to circumvent sequentiality and to ensure structural feasibility while keeping the computational process efficient and applicable. Thus, it is the objective to forge links between the single design stages to be able to exchange information and to obtain a better overall aircraft while decreasing the computational costs. Beyond that, some disciplines can not be computed/ used in an optimisation on a high-fidelity model due to immense computational costs. Examples include control, aeroservoelasticity, and post-buckling. Thus, it is aspirable to be able to compute these constraints on the low-fidelity model while still other disciplines are updated by the high-fidelity one. The idea is that information from a low-fidelity (beam-based) and a high-fidelity (shell-based) model are used by employing a multi-fidelity approaches. The envisioned design cycle is depicted in Figure \ref{fig:lifecycle}, comparing the past with the future.
Currently, while gaining knowledge about the design with an ongoing design process, the design freedom is decreasing. For conventional designs the lack of knowledge within the initial design stages is compensated by the use of empirical knowledge. However, for novel aircraft configurations this empirical knowledge is lack and needs to be replaced by physics-based knowledge, further motivating the need for incorporating high-fidelity information as early as possible.\\
This work is structured as follows. First, multi-fidelity optimisation is introduced with a brief overview of the current state of the art. Next, the two models used in this work are presented, i.e. a beam-based and a shell-based model. The beam-based model will hereafter be referred to as the low-fidelity (LF) model, while the shell-based model will be referred to as the high-fidelity (HF) model. As stated earlier, especially in large aerospace projects, multiple models often exist and transit through the different design phases. These models are not necessarily derived from each other, as they may serve different purposes. A beam-based model does not require detailed input from a three-dimensional computer-aided design (CAD) model and can be used to predict the overall elastic and inertial behaviour of slender structures. However such beam-based models are limited in predicting failure modes that involve displacement fields that are more complex than what traditional beam formulations are capable of predicting, such as stress concentrations or distortions in cross-sections caused by local buckling or any other local phenomenon. Shell-based models are often developed once the CAD geometry is available and can be employed to produce a thin-walled representation of the 3D structure. Here, under the assumption of plane stresses, more detailed failure modes can be captured under the additional computational cost involved with orders of magnitude more degrees-of-freedom, when compared to the beam-based model.\\
In this study, we present an investigation comparing these two models and exploring their application in multi-fidelity design optimisation. After aligning the two models in terms of geometry, the results compare the responses to different structural and aeroelastic analyses to gain an understanding of their behaviour for later use in multi-fidelity optimisation. Following this, existing methodologies for multi-fidelity design optimisation are reviewed, and finally, some concluding remarks are presented.

\section{Multi-Fidelity Multi-Disciplinary Design Optimisation}\label{ch:stateoftheart}
As introduced in the previous chapter, multi-fidelity design optimisation can significantly impact the design process of aerospace structures. In the following section, we briefly define the terminology of multi-fidelity and place it within the context of MDO (Multidisciplinary Design Optimisation). Generally speaking, the fundamental purpose of a model is to describe the relationship between an input and an output. In this work, we denote a model as a function $\mathcal{M}: \mathcal{X} \to \mathcal{Y}$, mapping an input $\boldsymbol{x} \in \mathcal{X} \subset \mathbb{R}^d$ from the design space, where $d \in \mathbb{N}$, to an output $\boldsymbol{y} \in \mathcal{Y} \subset \mathbb{R}^g$, where $g \in \mathbb{N}$. Typically, this relationship is unveiled by solving a Partial or Ordinary Differential Equation. Often, multiple models describe the same physical system, each with different approximation accuracies and computational costs. High-fidelity (HF) models are typically defined as having high accuracy but being costly to evaluate. Conversely, low-fidelity (LF) models are less accurate but fast to evaluate \cite{peherstorfer2018}. As mentioned in Section \ref{ch:intro}, some quantities or constraints are too expensive to compute using the HF model but need to be respected concurrently during its optimisation.\\
Anyway, optimisation falls into the realm of many-query applications, demanding repeated model evaluations to find the optimal solution. Consequently, relying solely on HF models often becomes impractical due to their substantial computational demands. Conversely, performing optimisation solely based on the LF model accelerates computations. However, due to the lower and usually unknown approximation accuracy, this may lead to unsatisfactory results, such as infeasible or suboptimal design points.\\
This motivates the development of multi-fidelity (MF) methods. These methods leverage the HF model to maintain accuracy while using the LF model, which captures the overall trend of the model output but may miss local details, to speed up computations. The goal is to obtain an optimisation result that is close to the outcome of a purely HF optimisation. This prompts the important question of how to manage these different models effectively.\\
A first and relatively simple approach is what the authors in \cite{2024Li-MFsurvey} call fixed methods. Fixed methods have a predetermined pattern of LF and HF model evaluations. However, this approach requires a good correlation between the models in use. For problems with low correlation, convergence issues can arise. In other words, the optimiser may be distracted by a sudden evaluation of the HF model after several iterations of the LF model. Another multi-fidelity approach frequently used in the literature is to sequentially increase the level of fidelity, indicated by a predefined indicator, as proposed by \cite{2022Wu}. The authors show that this approach can indeed decrease computational costs. These types of approaches are usually very economical since no further steps are needed to adjust the models. However, the aforementioned sequentiality remains present. To enhance the performance of multi-fidelity design optimisation, more sophisticated multi-fidelity approaches are required to be able to break out of the sequentiality.\\
\cite{peherstorfer2018} introduce three types of model management strategies aimed at balancing and correlating LF and HF model information to ensure accuracy while minimising computational cost. These three types are: \textit{Adaptation} \cite{2004Eldred, 2015Peherstorfer}, \textit{Fusion} \cite{Forrester2007}, and \textit{Filtering} \cite{2014Narayan}.\\
In addition to distinguishing between different strategies for managing MF models, LF models can be categorised based on their derivation. \cite{peherstorfer2018} define three different types of models: \textit{Simplified models}, which use expertise from the HF model to derive an independent model with lower fidelity and differing computational costs; \textit{Projection-based models}, which project the governing equations onto a low-dimensional subspace to yield a reduced model, exploiting the problem structure mathematically instead of relying on knowledge to derive a LF model; and \textit{Data-fit models}, which are derived from input and output data, functioning as a black box.\\
Let in the following $\mathcal{M}_{lf}$ be the LF and $\mathcal{M}_{hf}$ the HF model. The mapping from the design space $\mathcal{X}$ to the outputs can be written as 
\begin{equation}
	\begin{aligned}\label{eq:mfmodels}
		&\mathcal{M}_{lf}: \mathcal{X} \to \{ f_{lf}, \textbf{c}_{lf}, \nabla f_{lf},  \nabla \textbf{c}_{lf}\} \\
		&\mathcal{M}_{hf}: \mathcal{X} \to \{ f_{hf}, \textbf{c}_{hf}, \nabla f_{hf},  \nabla \textbf{c}_{hf} \}
	\end{aligned}
\end{equation}
where $f(\textbf{x}) \in \mathbb{R}, \nabla f(\textbf{x}) \in \mathbb{R}^{d}$ is the objective function and its gradients and $\textbf{c}(\textbf{x}) \in \mathbb{R}^{g}, \nabla \textbf{c}(\textbf{x}) \in \mathbb{R}^{g \times d}$ the constraints and their gradients. \\
One of the earliest introductions to using multiple models within a concurrent local optimisation process can be found in the works of \cite{Alexandrov1998, Alexandrov2001}. The authors present the so-called Trust Region Model Management (TRMM) strategy, which manages different numerical models based on the Trust Region approach. The standard Trust Region approach typically uses a quadratic approximation of the HF model, which is then optimised within the Trust Region. The authors in \cite{Alexandrov1998} generalise this idea by using the LF model instead, while occasionally evaluating the HF model to check progress.\\
Let $f_{hf}(\textbf{x}i)$ be the objective function of the HF model and $f{lf}(\textbf{x}_i)$ the LF model at the $i$-th design point $\textbf{x}_i$. If the LF model is constructed to satisfy first-order consistency, written as
\begin{equation}\label{eq:firstorderconsistency}
	\begin{split}
		f^{(i)}_{lf}(\textbf{x}_i) &= f^{(i)}_{hf}(\textbf{x}_i), \\
		\nabla f^{(i)}_{lf}(\textbf{x}_i) &= \nabla f^{(i)}_{hf}(\textbf{x}_i)
	\end{split}
\end{equation}
at every point $\textbf{x}_i$ of the $i$-th iteration. Subsequently, the optimisation problem inside the trust region is solved to find the next update point. Depending on a comparison between the LF and HF models, the step size is adjusted. This approach, due to its first-order consistency, provably converges to an HF local optimum. Following this, the method is extended to maximise the use of the LF model to further reduce computational cost by presenting three nonlinear programming algorithms and demonstrating their applicability on an Euler analysis with varying degrees of mesh refinement \cite{Alexandrov2001}.\\
Further works have built on this foundation. For instance, TRMM has been used with Bayesian model calibration to circumvent the need for HF derivatives \cite{March2012NoHF} and has introduced a modified Trust Region method called Trust Region Filter SQP \cite{2002fletcher, Elham2017}. To address differing numbers of design variables in LF and HF models, \cite{Robinson2008} apply a space mapping approach (first introduced by \cite{Brandler1994}) in combination with TRMM, mapping LF variables to HF ones. \cite{Leifsson2010} then applied the space mapping approach to transonic airfoil aerodynamics.\\
\cite{Bryson2017} introduced the multi-fidelity Broyden–Fletcher–Goldfarb–Shanno (MF-BFGS) optimisation algorithm, based on a quasi-Newton solver. This method demonstrated faster convergence and reduced computational costs compared to TRMM when applied to an unconstrained problem \cite{Bryson2017}. The MF-BFGS method was then used to solve an aeroelastic design optimisation problem, optimising the range of an aircraft using seven design variables \cite{2018Bryson}. Subsequently, \cite{Thelen2022} investigated the scalability of the MF-BFGS approach.\\
However, all the aforementioned methods assume an unconstrained optimisation problem, where constraints are introduced via penalties added to the objective function to address this limitation. Penalty methods only find approximate solutions. However, when combined with gradient-based methods, they exhibit numerical issues \cite{martins_engineering_2021}. Explicitly handling constraints, as done in Sequential Quadratic Programming (SQP) or within algorithms such as the Globally Converging Method of Moving Asymptotes (GCMMA) \cite{Svanberg1987}, has not yet been addressed in the literature.\\
As pointed out by \cite{2024Li-MFsurvey}, there is a lack of scalable methods that can handle high-dimensional inputs and outputs (large-scale constraints) without relying on aggregation methods, while efficiently leveraging gradient information. Engineering problems like the one introduced in Section \ref{ch:intro} are often constrained in nature, and therefore it is expected that an accelerated convergence would be achieved by developing a novel multi-fidelity optimisation method that accounts for an explicit handling of constraints, which is the ultimate goal of this line of research. Furthermore, a multi-fidelity scheme that accounts for a subset of constraints, available only in one model, shall also be considered.

\section{Aeroelastic Tailoring: Models}\label{ch:models}
In line with prior discussions, multi-fidelity models usually stem from one another, as in \textit{projection-based models}. However, as highlighted in Section \ref{ch:stateoftheart}, LF models also exist that are derived based on knowledge about the physical system, known as \textit{simplified models}, without requiring complex 3D representations of the geometry before the models can be generated. This scenario commonly arises in large interdisciplinary aerospace projects, where diverse frameworks coexist and an increasing amount of information about the system becomes available throughout the various design phases. These models represent the same physical system but are derived from different formulations, typically aligning with the information available at various design phases. Each model has its own strengths and weaknesses. Efficiently handling and exploiting these diverse capabilities and the generated data, however, presents a significant challenge in today's aerospace industry. To leverage multi-fidelity optimisation with these independent models, this work aims to bridge this gap for the use in Aeroelastic Tailoring.\\
Hereinafter, as previously introduced, two independent models are used. The HF model is a shell-based Nastran model, referred to as the Embraer Benchmark Wing, which serves as the ground truth. This model will be explained in more detail in Subsection \ref{ch:ebmw}.\\
The LF model, however, is a beam-based model that is part of the low-fidelity design framework \textit{Proteus}, developed by the Aerospace Structures and Materials department at Delft University of Technology, initiated with the work of \cite{de_breuker_energy-based_2011}, as described in Subsection \ref{ch:proteus}.\\
The overall optimisation problem is noted in Equation \ref{eq:generaloptimisationproblem}. The design variables are the lamination parameters (denoted with the superscript $l$) of each panel as well as their thickness, denoted with the superscript $t$.
\begin{equation}\label{eq:desvars}
	\textbf{x} = \biggl\{ \textbf{x}^{l}_1, x^t_1, ...,  \textbf{x}^{l}_{n_p}, x^t_{n_p} \biggr\}.
\end{equation}
Note that in this work, the lamination parameters as part of the design variables are denoted by $\textbf{x}^{l}_i$, whereas in the composite community they are often referred to as $\xi_{1,2,3,4}^{A,D}$. In the following subsections, both models are briefly explained.

\subsection{\textsc{Embraer} Benchmark Wing: A High-Fidelity Aeroelastic Tailoring Model  in Nastran}\label{ch:ebmw}
\begin{figure}[ht]
	\centering
	\includegraphics[width=12cm]{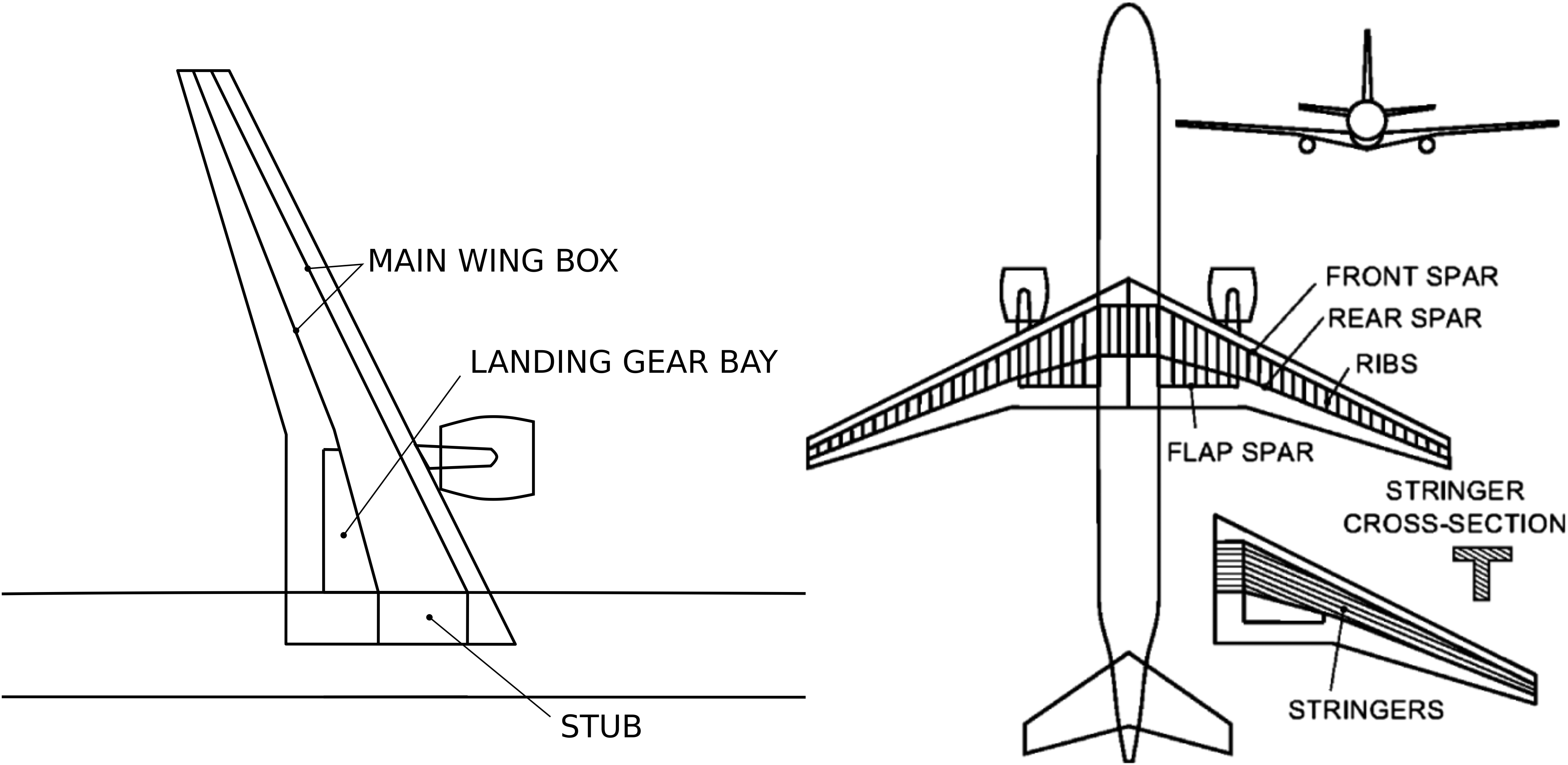}
	\caption{Aerodynamic planform and structural layout (from \cite{2019Silva}).}
	\label{fig:layout}
\end{figure}

The HF model is the  \textsc{Embraer} Benchmark Wing in \textsc{Nastran}. This model was developed to represent a generic aircraft, similar to the \textsc{Embraer} E-Jet family. It serves as a benchmark model for testing new analysis and design frameworks within an international Aeroelastic Tailoring consortium. In this model, the skin, spar, and rib covers are represented by shell elements, for which the material properties are specified separately via the matrices $\textbf{A}(\textbf{x})$ and $\textbf{D}(\textbf{x})$. The lamination parameters can be used with classical laminate theory to construct the following relationship:
\begin{equation}\label{eq:clt}
	\begin{bmatrix} N \\ M \end{bmatrix} =\begin{bmatrix} \textbf{A}(\textbf{x}) & \textbf{B}(\textbf{x}) \\ \textbf{B}(\textbf{x}) & \textbf{D}(\textbf{x}) \end{bmatrix} \begin{bmatrix} \epsilon^0 \\ \kappa \end{bmatrix}.
\end{equation}
The complete derivation of how to compute the matrix entries of $\textbf{A}(\textbf{x})$ and $\textbf{D}(\textbf{x})$ can be found in \cite{werter_aeroelastic_2017, 2019Silva}. These lamination parameters then define the material properties in the \textsc{MAT2} card. Furthermore, additional parts such as pylon/engine and aileron assemblies are modelled using flexible beam and shell elements to obtain a realistic representation. The aileron and other moving surfaces are likewise modelled via composite materials, similar to the main wingbox. Additionally, a realistic actuator system combined with a hinge mechanism is used. To ensure proper inertia properties of the system, the engine itself is modelled using rigid elements connected to the engine mount. Non-structural masses such as fuel tanks, high-lift devices, landing gear, accessories, and miscellaneous items are accounted for as lumped masses. This is crucial to maintain the correct dynamic behavior of the system. Similarly, the fuselage, payload, and tail are added as point masses and are used to obtain the proper trim condition.\\
The static and dynamic aerodynamic loads are computed using the Doublet Lattice Method (DLM) in \textsc{Nastran}. For further details, the reader is referred to \cite{2019Silva}. As can be seen, the structure is much finer discretised, leading to an expectedly more accurate computation of field quantities compared to the beam model, where information is condensed to the beam nodes. Additionally, more details are available, as shown in Figure  \ref{fig:layout}. Of course, there is still room to further increase the fidelity of the model, for example, by using higher-fidelity aerodynamic solvers. However, it's important to note that this work focuses on preliminary aircraft design, where computationally efficient models are still needed for use in a multi-query context.

\begin{figure}[ht]
	\centering
	\includegraphics[width=9cm]{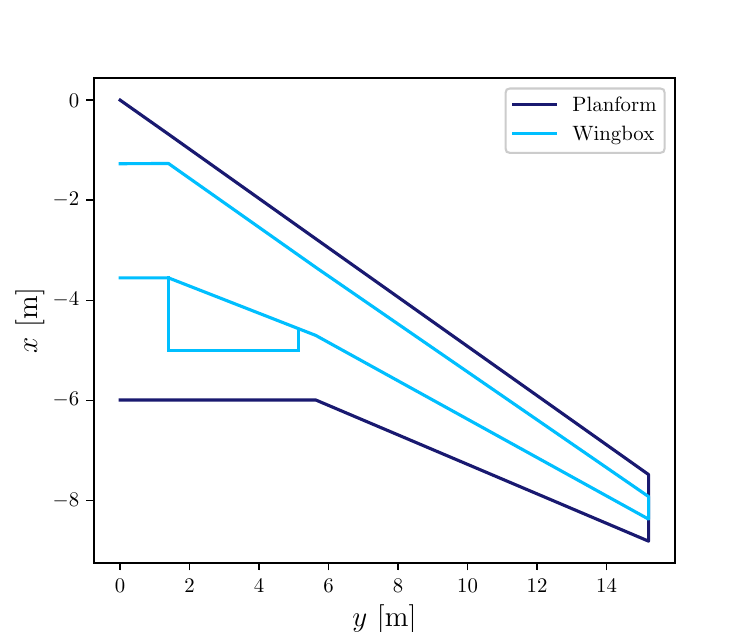}
	\caption{Planform measurements.}
	\label{fig:planform}
\end{figure}
\begin{figure}[ht]
	\centering
	\includegraphics[width=15cm]{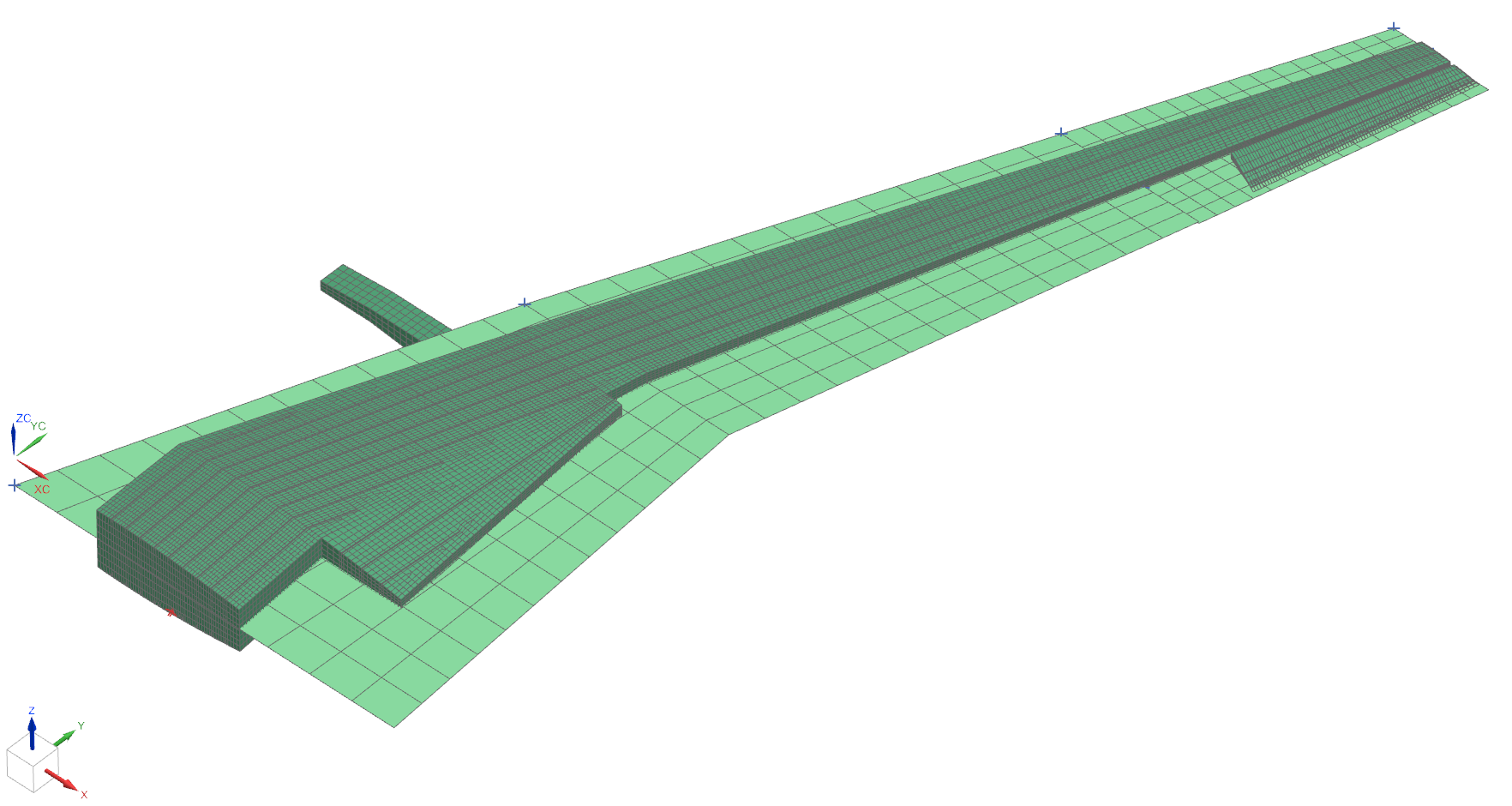}
	\caption{Shell-based model in \textsc{Nastran} including aerodynamic panels.}
	\label{fig:hfmodel}
\end{figure}
This model can then be used within \textsc{Nastran}'s optimisation environment \textsc{SOL200} to compute the corresponding constraints, arising for example from a buckling analysis with \textsc{SOL105} or a flutter analysis in \textsc{SOL145}.\\
This can be written as the input-output mapping, denoted as 
\begin{equation}
	\begin{aligned}
		\mathcal{M}_{hf}: \mathcal{X} \to \{ f_{hf}, \textbf{c}_{hf}, \nabla f_{hf},  \nabla \textbf{c}_{hf}\} \\
		f_{hf}(\textbf{x}) \in \mathbb{R} && \textbf{c}_{hf}(\textbf{x}) \in \mathbb{R}^g && \nabla f_{hf}(\textbf{x}) \in \mathbb{R}^d && \nabla \textbf{c}_{hf}(\textbf{x}) \in \mathbb{R}^{g \times d}
	\end{aligned}
\end{equation}

\subsection{\textsc{Proteus}: A Low-Fidelity Aeroelastic Tailoring Framework}\label{ch:proteus}
\begin{figure}[h]
	\centering
	\includegraphics[width=13cm]{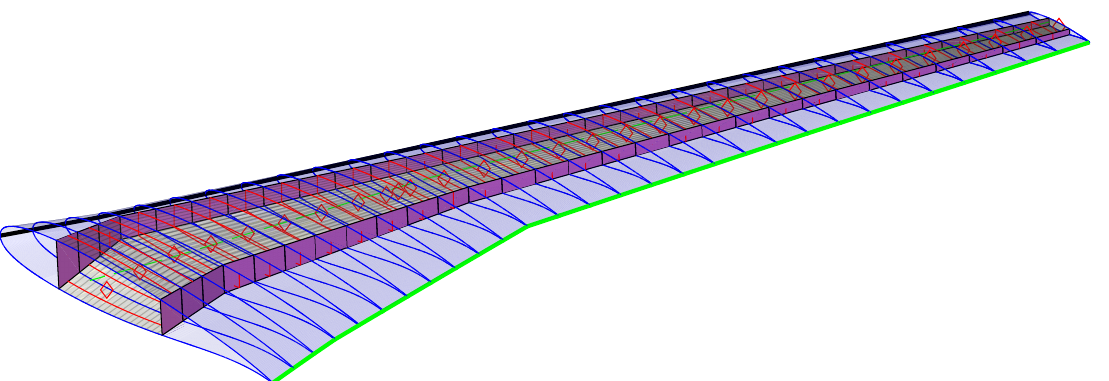}
	\caption{Beam representation of the wing structure. Diamonds represent the beam nodes. The beam nodes are chosen to be located at each rib, leading to 30 beam nodes and thus 29 beam elements.}
	\label{fig:beammodel}
\end{figure}

In general, the framework transforms a three-dimensional wingbox made of laminate panels into a nonlinear three-dimensional beam model. A panel in this setting can be, for instance, the upper and lower skin cover or the skin of the front and rear spar. Additionally, the wingbox can be discretised in the chord-wise direction with a user-defined number to increase the design freedom. The airfoil shape, as can be observed in \ref{fig:beammodel}, is used here to size the wingbox. However, the panel method used to compute the aerodynamic loads is based on a flat plate airfoil.
The lamination parameters can be used with the classical laminate theory to construct the relationship defined in Equation \ref{eq:clt}. This relationship encodes the dependency of the design variables with the stiffness of the system. A cross-section modeller \cite{ferede_cross-sectional_2014}, based on a variational approach, is used to obtain the element Timoshenko cross-sectional stiffness matrix $\textbf{C} \in \mathbb{R}^{6 \times 6}$ by relating the strains $\boldsymbol{\epsilon}$ to the applied forces and moment $\boldsymbol{\sigma}$ or $F_i,M_i$ respectively, as in 
\begin{equation}
	\boldsymbol{\sigma} = \textbf{C} \boldsymbol{\epsilon} \rightarrow \begin{bmatrix} F_1 & F_2 & F_3 & M_1 & M_2 & M_3 \end{bmatrix}^T =   \textbf{C} \begin{bmatrix} \epsilon_{11} & \epsilon_{12} & \epsilon_{13} & \kappa_{1} & \kappa_{2} & \kappa_{3} \end{bmatrix}^T
\end{equation}
with $\kappa_{1}$ as the twist and $\kappa_{2}, \kappa_{3}$ the bending curvatures. Therewith, the properties of a $2D$ cross section are mapped onto the corresponding beam node, leading to a $6 \times 6$ element Timoshenko beam stiffness matrix. The corresponding beam strain energy in the continuous form can be derived like 
\begin{align}
	\mathcal{U} = \frac{l_0}{2} \int_{0}^1  \boldsymbol{\epsilon}^{T} \textbf{C}  \boldsymbol{\epsilon} d\xi.
\end{align}
By discretisation and introduction of the element degrees of freedom $\textbf{p} \in \mathbb{R}^{n_{dof}}$, the linear beam element stiffness matrix, at this point neglecting geometric and material nonlinearities, can be computed by 
\begin{align}
\boldsymbol{\epsilon} = \textbf{B} \textbf{p} && \textbf{K}_{ij} = \frac{\partial^2 \mathcal{U}}{\partial \textbf{p}_i \partial \textbf{p}_j }
\end{align}
where the matrix $\textbf{B}$ interpolates the nodal quantities to strains $\boldsymbol{\epsilon}$. $n_{dof} \in \mathbb{N}$ is the number of degrees of freedom (DoF), leading to 12 DoF per beam element, thus $6$ DoF per node. The beam model is depicted in Figure \ref{fig:beammodel}, where diamonds represent the beam nodes. \\
Assuming a force vector $\textbf{f}$, the static solution is given as
\begin{equation}
	\textbf{K}(\textbf{p}) \textbf{p} = \textbf{f}.
\end{equation}
Note that geometrical nonlinearities are introduced by using the co-rotational framework of \cite{battini_co-rotational_2002}, decomposing large displacements and rotations into rigid body motion and small elastic deformations, ultimately leading to a dependence of the stiffness matrix $\textbf{K}$ on the displacements $\textbf{p}$. \\
After formulating the nonlinear structure, the aerodynamic forces and moments are computed via the continuous-time state-space unsteady vortex lattice method (UVLM), derived in \cite{2015Werter}, and mapped onto the structure. Assuming low to moderate subsonic Mach numbers, very high Reynolds numbers as well as small angles of attack, the flow can be assumed to be incompressible, inviscid and irrotational. These assumptions reduce the Navier-Stokes equations to the Laplace equation, written as 
\begin{equation}
	\nabla^2\Phi = 0 
\end{equation} 
To solve the problem, boundary conditions are applied, leading to 
\begin{align}
	(\nabla \Phi + \textbf{V}_\infty) \cdot \textbf{n} = 0 \ \text{on} \ \partial \Omega && \lim_{ | \textbf{x} - \textbf{x}_0 | \to \infty} \nabla \Phi = 0
\end{align} 
where $\Phi$ is the velocity potential, $\textbf{V}_\infty$ the free-stream velocity, $\textbf{n}$ the surface normal vector and $\partial \Omega$ the boundary (wing surface). \\
The resulting overall nonlinear aeroelastic system does not guarantee finding an equilibrium point right away. That motivates the need for an iterative solution to obtain the nonlinear static response. Starting with
\begin{equation}
	\textbf{f}_s(\textbf{p}) = \textbf{f}_{ext}(\textbf{p}),
\end{equation} 
where the subscript $s$ denotes the structural force, a predictor-corrector Newton-Raphson solver is used to solve 
\begin{equation}
	\left( \frac{\partial \textbf{f}}{\partial \textbf{p}} - \frac{\partial \lambda \textbf{f}_{ext}}{\partial \textbf{p}} \right) \delta \textbf{p} = \textbf{f} - \textbf{f}_{ext} = \textbf{R}.
\end{equation} 
The solution can then be used to compute the corresponding stresses which can be applied to, for instance, the Tsai-Wu failure criterion $\textbf{w}(\boldsymbol{\sigma})$ to assess the static strength of the structure. To reduce the number of constraints, only the 8 most critical Tsai-Wu strain factors per panel are considered \cite{ijsselmuiden_2011}, leading to
\begin{equation}\label{eq:c-tw}
	\textbf{c}_{tw} = \textbf{w}_{crit}(\boldsymbol{\sigma}) \leq 0. 
\end{equation} 
The buckling analysis is based on the assumptions that no global buckling can occur due to sufficient stability of stiffeners and ribs. By additionally computing the geometric stiffness matrix $\textbf{K}_g$, the buckling factor $\lambda_b$ can be found by solving the eigenvalue problem 
\begin{equation}
	\left( \textbf{K} + \lambda_b \textbf{K}_g \right) \textbf{a} = \textbf{0}.
\end{equation} 
To reduce the number of constraints, only the eight most critical eigenvalues per buckling analysis region (defined as a panel between two ribs and two stringers) are formulated as constraints, leading to
\begin{equation}\label{eq:c-b}
	\textbf{c}_{b} = -\boldsymbol{\lambda}_{b,crit} + 1 \leq 0.
\end{equation} 
To compute the aeroelastic stability of the system, the equilibrium of forces and moments of the internal structure $\textbf{f}_s$ and the externally applied loads must be regarded. The external forces are split up into applied aerodynamic loads $\textbf{f}_a$ and remaining external forces due to e.g. gravity or thrust $\textbf{f}_e$, given by 
\begin{equation} \label{eq:force-equil}
	\textbf{f}_s - \textbf{f}_{ext} = \textbf{f}_s - \textbf{f}_a - \textbf{f}_e = \textbf{0}
\end{equation} 
By linearising Equation \ref{eq:force-equil}, the corresponding stiffness matrices $\textbf{K}_a$, $\textbf{K}_e$ and $\textbf{K}_s$ can be obtained and the stability of this static aeroelastic equilibrium is governed by
\begin{equation}\label{eq:static_equil}
	\left( \lambda_s \textbf{K}_a + \textbf{K}_e - \textbf{K}_s \right) \Delta \textbf{p} = \textbf{0}.
\end{equation} 
The dynamic aeroelastic analysis is carried out by linearising the system around the static aerodynamic equilibrium and by using the state-space formulation for both the aerodynamic and the structural part. It should be mentioned that in this chapter many steps are left out for the sake of brevity. For more details the reader is referred to \cite{de_breuker_energy-based_2011, werter_aeroelastic_2017}. However, ultimately ending up at the well-known continuous-time state-space formulation of the aeroelastic system, written as
\begin{equation}
	\dot{\textbf{s}} = \textbf{A} \textbf{s} + \textbf{B} \boldsymbol{\alpha}_{air}
\end{equation} 
with $\boldsymbol{\alpha}_{air}$ as the perturbation angle of attack of the induced free stream flow, incorporating structural and aerodynamic forces and their coupling. For dynamic aeroelastic stability, and thus to prevent the structure from flutter, the eigenvalue problem for the state matrix $\textbf{A}$ must be solved once again, expressed as:
\begin{equation}\label{eq:stability_analysis}
	\left( \textbf{A} - \textbf{I} \lambda_f \right) \boldsymbol{\varphi} = \textbf{0}.
\end{equation} 
Anew, only the ten most critical eigenvalues $\lambda_{f,crit}$ are considered \cite{werter_aeroelastic_2017}, leading to
\begin{equation}\label{eq:c-ds}
	\textbf{c}_{ds} = \Re(\lambda_{f,crit}) \leq 0.
\end{equation} 
Furthermore, two more types of constraints are formulated. The Aileron Effectiveness is constrained as follows
\begin{align}\label{eq:c-ae}
	c_{ae} = \eta_{eff} - \eta_{eff,min} \leq 0,
\end{align}
to ensure a safe manoeuvrability of the aircraft as well as the Angle of Attack $\alpha$ is constrained by using an upper and lower bound, written as 
\begin{equation}
	\begin{aligned}\label{eq:c-aoa}
		\textbf{c}_{AoA,lb} &= -\alpha - \alpha_{lb} \leq 0, \\
		\textbf{c}_{AoA,ub} &= \alpha - \alpha_{ub} \leq 0
	\end{aligned} 
\end{equation}
adding two more constraints per aerodynamic cross-section. \\
Lastly, the lamination parameter feasibility constraints ensure that the laminates satisfy certain interdependencies and are analytic equations, derived in \cite{raju_further_2014}, ending up in 6 inequality constraints per panel.
\begin{equation}\label{eq:c-lf}
	\textbf{c}_{feas} = \begin{bmatrix} \textbf{g}_1^\top(\textbf{x}) \ \textbf{g}_2^\top(\textbf{x}) \ \textbf{g}_3^\top(\textbf{x}) \ \textbf{g}_4^\top(\textbf{x}) \ \textbf{g}_5^\top(\textbf{x}) \ \textbf{g}_6^\top(\textbf{x}) \end{bmatrix}^\top \leq \textbf{0}.
\end{equation}
Note that each of the six constraints is evaluated for all $n_p$ panels, thus $\textbf{g}_i: \mathcal{X} \to \mathbb{R}^{n_p}$ and $\textbf{c}_{feas}: \mathcal{X} \to \mathbb{R}^{6n_p}$.\\
Finally, the constraints can be concatenated to form together with the objective function $f(\textbf{x})$ the outputs of the model, written as $\textbf{c}_{lf}(\textbf{x}) = \{ \textbf{c}_{tw}, \textbf{c}_{b},\textbf{c}_{ds}, c_{ae}, \textbf{c}_{AoA}, \textbf{c}_{feas} \}^\top$. All categories of constraints besides the lamination parameter feasibility need to be taken into account per loadcase.\\
Advantages of Proteus are the efficiency in evaluating the model due to its low-fidelity formulation. As several authors showed before, the framework shows excellent results compared to models with higher fidelity \cite{werter_aeroelastic_2017,2020Natella,2021Rajpal}. This property is particularly important if, in the future, constraints arising from disciplines like control or aeroservoelasticity are to be used in preliminary design. Moreover, due to the availability of the source code and the use of analytical gradients, the framework can compute sensitivities of the outputs w.r.t. the inputs very efficiently. 
Thus, the overall LF model can be summarised in terms of the earlier introduced notation as 
\begin{equation}
	\begin{aligned}
		\mathcal{M}_{lf}: \mathcal{X} \to \{ f_{lf}, \textbf{c}_{lf}, \nabla f_{lf},  \nabla \textbf{c}_{lf}\} \\
		f_{lf}(\textbf{x}) \in \mathbb{R} && \textbf{c}_{lf}(\textbf{x}) \in \mathbb{R}^k && \nabla f_{lf}(\textbf{x}) \in \mathbb{R}^d && \nabla \textbf{c}_{lf}(\textbf{x}) \in \mathbb{R}^{k \times d}
	\end{aligned}
\end{equation}
Note that the number of constraints, denoted as $k$, might differ from that in the HF model, denoted as $g$. This distinction is made to underscore the possibility of constraints that are exclusively available in one model.

\section{Results}
This section is dedicated to comparing the models introduced in Section \ref{ch:models}. In the works cited thus far, the LF framework was typically compared against a model reconstructed in Nastran to ensure consistency between both models. However, as mentioned earlier, this work takes the HF model as the ground truth and rebuilds this model within the LF framework. Of particular interest is the lack of detail in the LF model compared to the HF one. The latter exhibits much more detail that cannot be represented in the simplified model, such as inspection holes.\\
To ensure that the two models behave similarly and enable their output information to be used in multi-fidelity optimisation, a comparison of both models is performed. However, it should be noted that inaccuracies may be observed in specific cases, which will later be addressed by the multi-fidelity framework. Additionally, while the HF model is an approximation of the real-world behaviour, its response is considered the ground truth in this study. 
After aligning the two models to represent the same geometry and design regions, multiple load cases/analysis are performed to gain insights into the accuracy of both: 
\begin{enumerate}
	\item A static linear analysis with out-of-plane tip load as well as a torsional moment applied at the tip to compare the stiffness properties of both models
	\item A structural modal analysis to validate the stiffness properties in combination with masses 
	\item A aeroelastic stability analysis to compare the structural properties in combination with the used panel codes (DLM in the HF model and UVLM in the LF model)
\end{enumerate}
It is worth mentioning that discrepancies between the two models due to non-represented details are expected and further motivate the use of multi-fidelity algorithms.  

\subsection{Case 1: Static Analysis}
\begin{figure}[h!]
	\centering
	\begin{subfigure}{.45\textwidth}
		\centering
		\includegraphics[width=\textwidth]{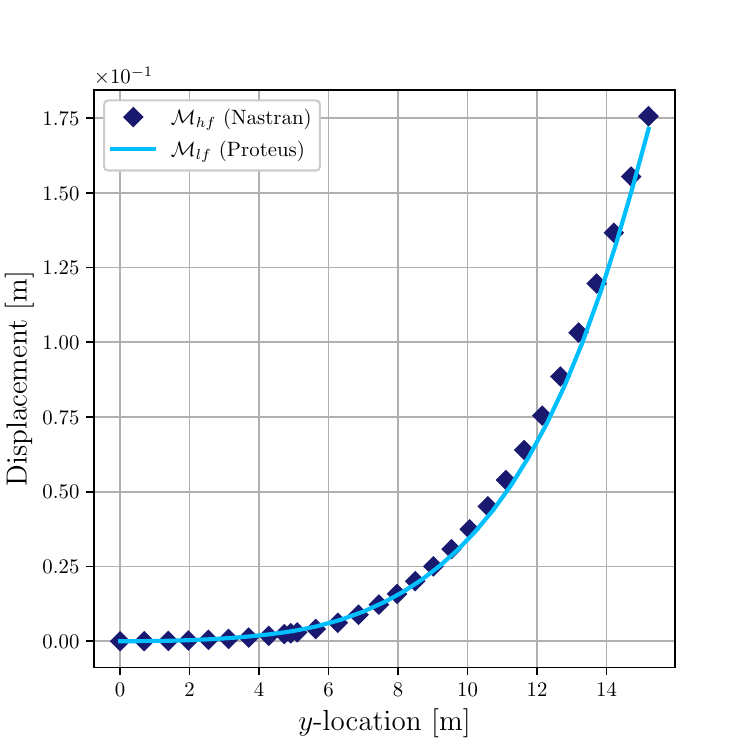}
		\caption{Out-of-plane tip load.}
		\label{fig:tip_load}
	\end{subfigure}
	\begin{subfigure}{.45\textwidth}
		\centering
		\includegraphics[width=\textwidth]{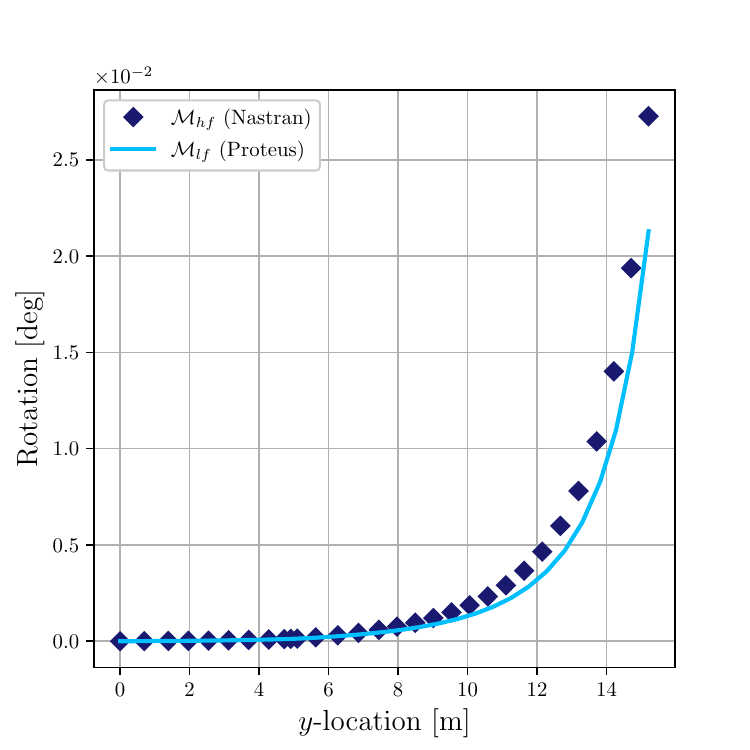}
		\caption{Tip torsion.}
		\label{fig:tip_torsion}
	\end{subfigure}
	\caption{Comparison of static load case.}
	\label{fig:static}
\end{figure}

\begin{figure}[ht]
	\centering
	\includegraphics[width=8cm]{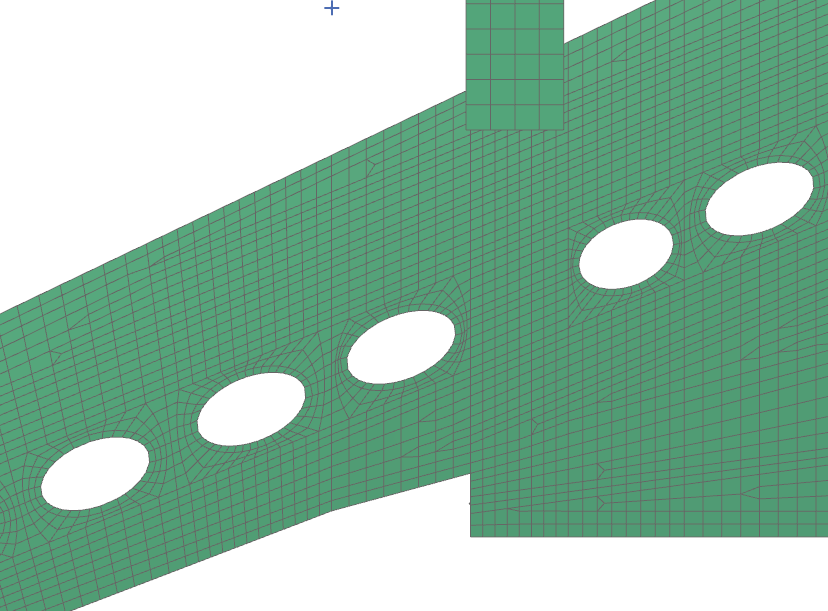}
	\caption{Inspection holes in the HF model.}
	\label{fig:inspectionholes}
\end{figure}

In this subsection, a linear static analysis is conducted on both models to compare their accuracy and stiffness. Two different loads are applied: an out-of-plane tip load ($F_z$), as illustrated in Figure \ref{fig:tip_load}, and a torsional load around the $y$-axis ($M_y$), depicted in Figure  \ref{fig:tip_torsion}. Assuming the HF model serves as the reference, the out-of-plane deflection of the LF model exhibits very good agreement with the HF model, with a relative error of only 5.33\%.\\
However, when examining the torsional load, the agreement between the two models is less accurate, with a relative error of 23.99\%. A probable explanation for this discrepancy lies in the presence of inspection holes in the HF model, as depicted in Figure \ref{fig:inspectionholes}, which are not present in the LF model. In the latter, the lower skin panels are assumed to be closed, resulting in a higher torsional stiffness.

\subsection{Case 2: Structural Modal Analysis}
\begin{figure}[h!]
	\centering
	\begin{subfigure}{.45\textwidth}
		\centering
		\includegraphics[width=\textwidth]{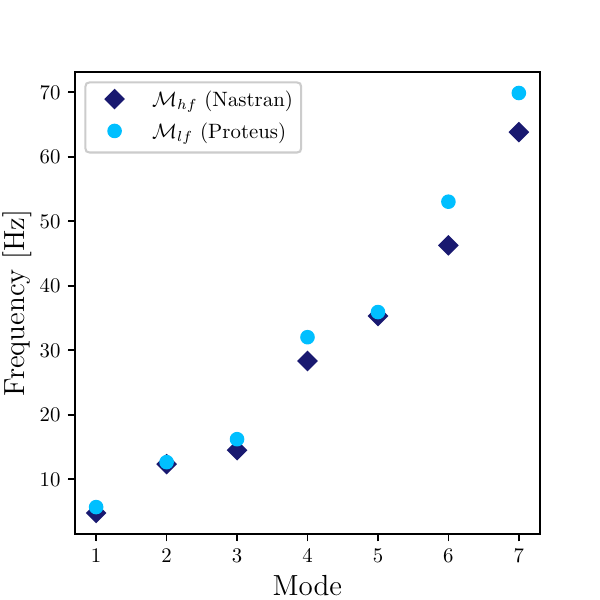}
		\caption{Comparison of eigenvalues.}
		\label{fig:eigfrequ_structure}
	\end{subfigure}
	\begin{subfigure}{.45\textwidth}
		\centering
		\includegraphics[width=\textwidth]{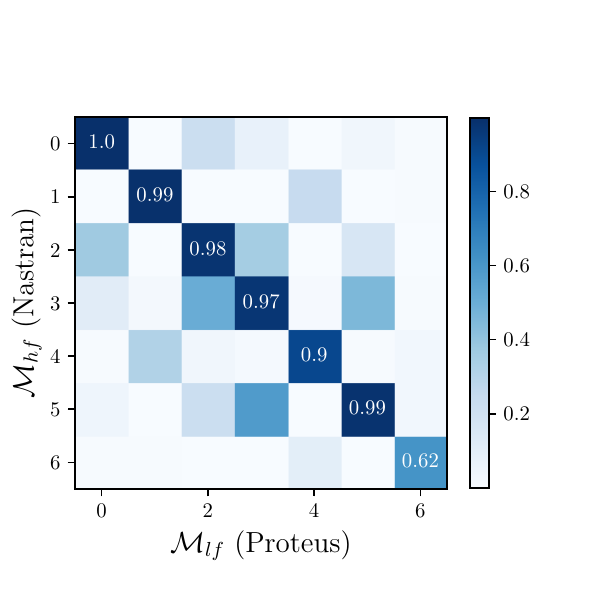}
		\caption{MAC of structural analysis.}
		\label{fig:mac_structure}
	\end{subfigure}
	\caption{Modal analysis of structure.}
	\label{fig:modal_analysis_structure}
\end{figure}

In this second subsection, a structural modal analysis is conducted to assess not only the agreement regarding the stiffness of the two models but also to consider the mass distributions. Figure \ref{fig:eigfrequ_structure} compares the computed eigenvalues of both models, revealing a very good agreement. Subsequently, the Mode Assurance Criterion (MAC) is utilised, expressed as:
\begin{equation}\label{eq:mac}
	\text{MAC}(\boldsymbol{\varphi}_i,\boldsymbol{\varphi}_j) = \frac{|\boldsymbol{\varphi}_i^*\boldsymbol{\varphi}_j|^2}{(\boldsymbol{\varphi}_i^*\boldsymbol{\varphi}_i) (\boldsymbol{\varphi}_j^* \boldsymbol{\varphi}_j)}
\end{equation}
with $\boldsymbol{\varphi}^*_{i}$ representing the complex conjugate transpose, which for real-based vectors becomes simply the transpose $\boldsymbol{\varphi}^*_{i}=\boldsymbol{\varphi}^\top_{i}$; where $\boldsymbol{\varphi}_i$ is the $i$-th eigenvector resulting from the following eigenvalue analysis:
\begin{equation}
(\textbf{K}-\omega^2\textbf{M}) \boldsymbol{\varphi} = \boldsymbol{0}.
\end{equation}
The mode shapes are compared in Figure \ref{fig:mac_structure}. A MAC value of $1$ indicates an exact agreement between the two mode shapes, while a value of $0$ suggests a mismatch. When comparing the two models, the main diagonal shows ones. As depicted in the aforementioned figure, the structural results once again demonstrate a very good agreement.

\subsection{Case 3: Aeroelastic Stability Analysis}
\begin{figure}[h!]
	\centering
	\begin{subfigure}{.45\textwidth}
		\centering
		\includegraphics[width=\textwidth]{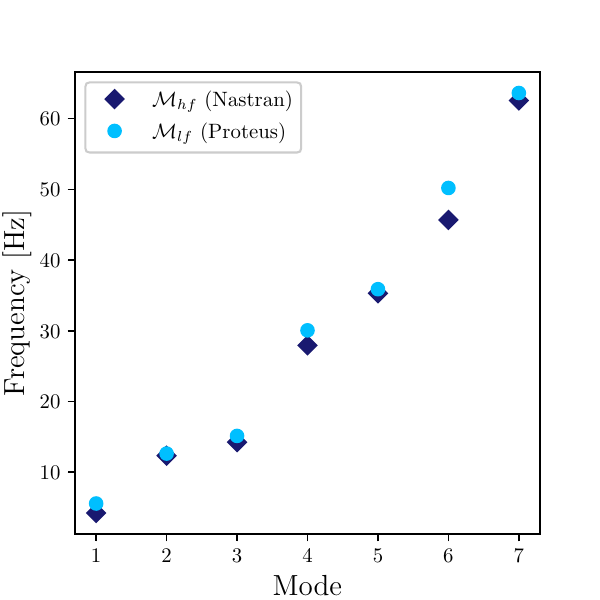}
		\caption{Comparison of eigenvalues.}
		\label{fig:eigfrequ_aeroelastic}
	\end{subfigure}
	\begin{subfigure}{.45\textwidth}
		\centering
		\includegraphics[width=\textwidth]{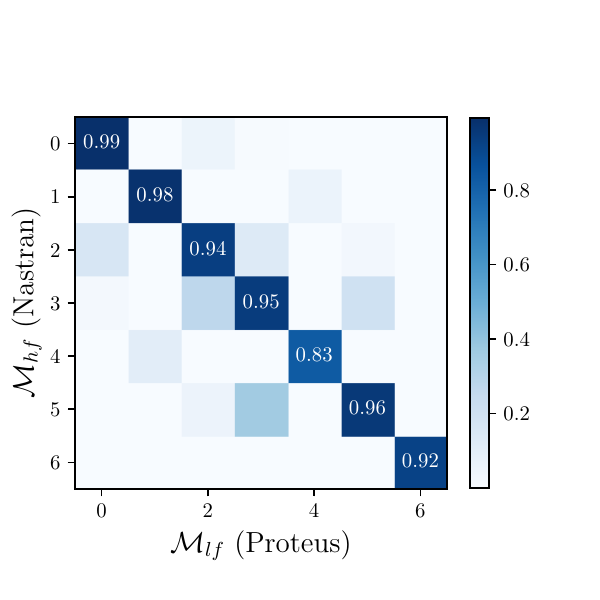}
		\caption{MAC of aeroelastic analysis.}
		\label{fig:mac_aeroelastic}
	\end{subfigure}
	\caption{Aeroelastic stability analysis at $V=140 \ \text{m/s}$ and $\text{Ma} = 0.69$.}
	\label{fig:modal_analysis_aeroelastic}
\end{figure}

The last case involves an aeroelastic stability analysis to incorporate the aerodynamics of the two models. However, no flutter analysis is performed; instead, an aeroelastic stability analysis is conducted at a given cruise speed of $V = 140 \ \text{m/s}$ and a given Mach number of $\text{Ma} = 0.69$. Since the LF model utilises a state-space formulation for the aeroelastic system, the eigenvalue analysis is performed at the specified speed, at which point the system is linearised. Therefore, Equation \ref{eq:stability_analysis} is solved to obtain the eigenvalues and the corresponding eigenvectors. Once again, MAC values are computed, see Equation \ref{eq:mac}. However, the eigenvectors of the aeroelastic system are now complex.  Figure \ref{fig:eigfrequ_aeroelastic} illustrates the comparison of the eigenfrequencies of the aeroelastic system, once again demonstrating very good agreement. Similarly, the mode shapes exhibit a good accordance, with values close to one along the main diagonal of the MAC matrix.

\section{Conclusion}
Within this section, a brief summary as well as an outlook for further research is given.

\subsection{Summary}
This work aims to advance multi-fidelity optimisation in preliminary aircraft design, targeting a two-sided goal. Firstly, it seeks to integrate high-fidelity information earlier in the design process. By replacing costly high-fidelity with inexpensive low-fidelity model evaluations, this line of research aims to leverage these insights for concurrent design optimisation.\\
Typically, models in aircraft design are derived from one another. However, it is often the case that different frameworks exist, with one model representing more detail than the other. The literature on multi-fidelity is first presented to provide a foundation. Following this, two models intended for use in a multi-fidelity context are introduced.\\
To ensure both models reflect the same trends, various analyses are performed to validate the approach's applicability. The work demonstrates that the low-fidelity framework \textsc{Proteus} can efficiently remodel existing complex wing structures with high accuracy, making it suitable for the envisioned multi-fidelity framework. Additionally, the low-fidelity model can compute gradients cheaply, which can be combined with high-fidelity (HF) model information in a concurrent design optimisation framework.\\
At specific design points, both models can be evaluated, yielding constraints at both LF and HF levels, along with their gradients. This capability enables a robust and efficient optimisation process, harnessing the strengths of both fidelity levels. The underlying concept is to learn and integrate the differences in accuracy between the models and the level of detail they capture into the design process. For example, consider aeroelastic stability, where the real part of the eigenvalue serves as a constraint. This constraint can be computed by both the HF and LF models, albeit with some approximation error. This error is then mitigated by a function learned from the model discrepancies, which can subsequently be utilised in the optimisation process.\\
Moreover, another compelling scenario arises when there are constraints that are too computationally expensive to be evaluated with a HF model. Despite this, these constraints are still accounted for, originating solely from the LF model.

\subsection{Outlook}
After exploring the applicability of the LF model alongside the HF model, further research aims to devise a suitable gradient-based algorithm for this type of multi-fidelity problem. The considerations outlined above necessitate a certain adaptability of the optimisation algorithm. As highlighted in Section \ref{ch:stateoftheart}, existing algorithms typically rely on unconstrained approaches, incorporating constraints through penalty terms (see, for example, \cite{Bryson2017}). However, as observed by \cite{martins_engineering_2021}, representing constraints using penalties can lead to convergence issues in gradient-based solution methods. Additionally, authors often resort to constraint aggregation methods to reduce the number of constraints, which further approximates the solution. \cite{Thelen2022} mentions that constraint aggregation, together with mesh regeneration, may lead to noisy responses that can slow convergence. \\
In summary, there is a pressing need for multi-fidelity schemes that explicitly address constraints while being scalable to higher-dimensional problems with large-scale constraints, such as those encountered in Aeroelastic Tailoring and many other MDO problems. This algorithm must effectively utilise available information—objective values, constraint values, and their derivatives—while modelling the correlation between the two fidelity levels. Methods like SQP or GCMMA \cite{Svanberg1987}, which explicitly handle constraints, are slated for further development in a multi-fidelity context. This development will be complemented by an adaptation approach, as discussed in Section \ref{ch:stateoftheart}, where the discrepancy between the two fidelities is modelled using a surrogate. Since this approach becomes computationally expensive when dealing with thousands of constraints, the error may be described on a lower-dimensional submanifold, akin to the approach outlined in \cite{2024Maathuis}. However, unlike the aforementioned approach, since gradients are available, Active Subspace methods can be employed to compute the basis vectors.

\section{Contact Author Email Address}
mailto: h.f.maathuis@tudelft.nl

\section{Copyright Statement}
\begin{small}
	The authors confirm that they, and/or their company or organisation, hold copyright on all of the original material included in this paper. The authors also confirm that they have obtained permission, from the copyright holder of any third party material included in this paper, to publish it as part of their paper. The authors confirm that they give permission, or have obtained permission from the copyright holder of this paper, for the publication and distribution of this paper as part of the ICAS proceedings or as individual off-prints from the proceedings.
\end{small}

\section*{Acknowledgement}
This research was supported by the ATED project in collaboration with \textsc{Embraer S.A.}. Special thanks to Alex Pereira do Prado and Pedro Higino Cabral for their invaluable support. Additionally, the authors thank our colleague Xavi Carrillo C\'orcoles for his assistance with \textsc{Proteus} and \textsc{Nastran}.

\bibliographystyle{nar}   
\bibliography{icas-bib.bib} 

\begin{thebibliography}{10}

\bibitem{1986Shirk}
Shirk, M.~H., Hertz, T.~J., and Weisshaar, T.~A. (January, 1986)
Aeroelastic tailoring - {Theory}, practice, and promise.
{\em Journal of Aircraft,} {\bf 23}(1), 6--18.

\bibitem{2013Dillinger}
Dillinger, J. K.~S., Klimmek, T., Abdalla, M.~M., and Gürdal, Z. (July, 2013)
Stiffness {Optimization} of {Composite} {Wings} with {Aeroelastic}
  {Constraints}.
{\em Journal of Aircraft,} {\bf 50}(4), 1159--1168.

\bibitem{1998Mavris}
Mavris, D., DeLaurentis, D., Bandte, O., and Hale, M. (January, 1998)
A stochastic approach to multi-disciplinary aircraft analysis and design.

\bibitem{Kennedy2014}
Kennedy, G.~J. and Martins, J. R. R.~A. (December, 2014)
A parallel aerostructural optimization framework for aircraft design studies.
{\em Structural and Multidisciplinary Optimization,} {\bf 50}(6), 1079--1101.

\bibitem{Kennedy2014-2}
Kenway, G. K.~W. and Martins, J. R. R.~A. (January, 2014)
Multipoint {High}-{Fidelity} {Aerostructural} {Optimization} of a {Transport}
  {Aircraft} {Configuration}.
{\em Journal of Aircraft,} {\bf 51}(1), 144--160.

\bibitem{peherstorfer2018}
Peherstorfer, B., Willcox, K., and Gunzburger, M. (June, 2018)
Survey of multifidelity methods in uncertainty propagation, inference, and
  optimization.
arXiv:1806.10761 [cs, math, stat].

\bibitem{2024Li-MFsurvey}
Li, K. and Li, F. (February, 2024)
Multi-{Fidelity} {Methods} for {Optimization}: {A} {Survey}.
arXiv:2402.09638 [cs].

\bibitem{2022Wu}
Wu, N., Mader, C.~A., and Martins, J. R. R.~A. (April, 2022)
A {Gradient}-based {Sequential} {Multifidelity} {Approach} to
  {Multidisciplinary} {Design} {Optimization}.
{\em Structural and Multidisciplinary Optimization,} {\bf 65}(4), 131.

\bibitem{2004Eldred}
Eldred, M., Giunta, A., and Collis, S. (August, 2004)
Second-{Order} {Corrections} for {Surrogate}-{Based} {Optimization} with
  {Model} {Hierarchies}.

\bibitem{2015Peherstorfer}
Peherstorfer, B. and Willcox, K. (July, 2015)
Dynamic data-driven reduced-order models.
{\em Computer Methods in Applied Mechanics and Engineering,} {\bf 291}, 21--41.

\bibitem{Forrester2007}
Forrester, A.~I., Sóbester, A., and Keane, A.~J. (December, 2007)
Multi-fidelity optimization via surrogate modelling.
{\em Proceedings of the Royal Society A: Mathematical, Physical and Engineering
  Sciences,} {\bf 463}(2088), 3251--3269.

\bibitem{2014Narayan}
Narayan, A., Gittelson, C., and Xiu, D. (January, 2014)
A {Stochastic} {Collocation} {Algorithm} with {Multifidelity} {Models}.
{\em SIAM Journal on Scientific Computing,} {\bf 36}(2), A495--A521.

\bibitem{Alexandrov1998}
Alexandrov, N.~M., Dennis, J.~E., Lewis, R.~M., and Torczon, V.
A trust-region framework for managing the use of approximation models in
  optimization. (February, 1998).

\bibitem{Alexandrov2001}
Alexandrov, N.~M., Lewis, R.~M., Gumbert, C.~R., Green, L.~L., and Newman,
  P.~A. (November, 2001)
Approximation and {Model} {Management} in {Aerodynamic} {Optimization} with
  {Variable}-{Fidelity} {Models}.
{\em Journal of Aircraft,} {\bf 38}(6), 1093--1101.

\bibitem{March2012NoHF}
March, A. and Willcox, K. (May, 2012)
Provably {Convergent} {Multifidelity} {Optimization} {Algorithm} {Not}
  {Requiring} {High}-{Fidelity} {Derivatives}.
{\em AIAA Journal,} {\bf 50}(5), 1079--1089.

\bibitem{2002fletcher}
Fletcher, R., Gould, N. I.~M., Leyffer, S., Toint, P.~L., and Wächter, A.
  (January, 2002)
Global {Convergence} of a {Trust}-{Region} {SQP}-{Filter} {Algorithm} for
  {General} {Nonlinear} {Programming}.
{\em SIAM Journal on Optimization,} {\bf 13}(3), 635--659.

\bibitem{Elham2017}
Elham, A. and Van~Tooren, M. J.~L. (May, 2017)
Multi-fidelity wing aerostructural optimization using a trust region
  filter-{SQP} algorithm.
{\em Structural and Multidisciplinary Optimization,} {\bf 55}(5), 1773--1786.

\bibitem{Robinson2008}
Robinson, T.~D., Eldred, M.~S., Willcox, K.~E., and Haimes, R. (November, 2008)
Surrogate-{Based} {Optimization} {Using} {Multifidelity} {Models} with
  {Variable} {Parameterization} and {Corrected} {Space} {Mapping}.
{\em AIAA Journal,} {\bf 46}(11), 2814--2822.

\bibitem{Brandler1994}
Bandler, J., Biernacki, R., {Shao Hua Chen}, Grobelny, P., and Hemmers, R.
  (December, 1994)
Space mapping technique for electromagnetic optimization.
{\em IEEE Transactions on Microwave Theory and Techniques,} {\bf 42}(12),
  2536--2544.

\bibitem{Leifsson2010}
Leifsson, L. and Koziel, S. (May, 2010)
Multi-fidelity design optimization of transonic airfoils using shape-preserving
  response prediction.
Vol.~1,  pp. 1311--1320.

\bibitem{Bryson2017}
Bryson, D. and Rumpfkeil, M.~P. (January, 2017)
Comparison of {Unifed} and {Sequential}-{Approximate} {Approaches} to
  {Multifdelity} {Optimization}.
In \emph{58th {AIAA}/{ASCE}/{AHS}/{ASC} {Structures}, {Structural} {Dynamics},
  and {Materials} {Conference}} Grapevine, Texas: American Institute of
  Aeronautics and Astronautics.

\bibitem{2018Bryson}
Bryson, D. and Rumpfkeil, M.
Aeroelastic {Design} {Optimization} using a {Multifidelity} {Quasi}-{Newton}
  {Method}. (January, 2018).

\bibitem{Thelen2022}
Thelen, A.~S., Bryson, D.~E., Stanford, B.~K., and Beran, P.~S. (April, 2022)
Multi-{Fidelity} {Gradient}-{Based} {Optimization} for {High}-{Dimensional}
  {Aeroelastic} {Configurations}.
{\em Algorithms,} {\bf 15}(4), 131.

\bibitem{martins_engineering_2021}
Martins, J. R. R.~A. and Ning, A. (November, 2021)
Engineering {Design} {Optimization},
Cambridge University Press,  1 edition.

\bibitem{Svanberg1987}
Svanberg, K.
The method of moving asymptotes—a new method for structural optimization.
  (February, 1987).

\bibitem{de_breuker_energy-based_2011}
De~Breuker, R.
Energy-based aeroelastic analysis and optimisation of morphing wings
PhD thesis s.n. S.l. (2011)
OCLC: 840445505.

\bibitem{werter_aeroelastic_2017}
Werter, N.
Aeroelastic {Modelling} and {Design} of {Aeroelastically} {Tailored} and
  {Morphing} {Wings}
PhD thesis Delft University of Technology (2017).

\bibitem{2019Silva}
Silve, G., Prado, A., Cabral, P., De~Breuker, R., and Dillinger, J. (2019)
Tailoring of a Composite Regional Jet Wing Using the Slice and Swap Method.

\bibitem{ferede_cross-sectional_2014}
Ferede, E. and Abdalla, M. (January, 2014)
Cross-sectional modelling of thin-walled composite beams.
In \emph{55th {AIAA}/{ASME}/{ASCE}/{AHS}/{ASC} {Structures}, {Structural}
  {Dynamics}, and {Materials} {Conference}} National Harbor, Maryland: American
  Institute of Aeronautics and Astronautics.

\bibitem{battini_co-rotational_2002}
Battini, J.-M. and Pacoste, C. (February, 2002)
Co-rotational beam elements with warping effects in instability problems.
{\em Computer Methods in Applied Mechanics and Engineering,} {\bf 191}(17-18),
  1755--1789.

\bibitem{2015Werter}
Werter, N. P.~M., De~Breuker, R., and Abdalla, M.~M.
Continuous-{Time} {State}-{Space} {Unsteady} {Aerodynamic} {Modeling} for
  {Efficient} {Loads} {Analysis}. (March, 2018).

\bibitem{ijsselmuiden_2011}
Ijsselmuiden, S.~T.
Optimal Design of Variable Stiffness Composite Structures Using Lamination
  Parameters
PhD thesis Delft University of Technology (2011).

\bibitem{raju_further_2014}
Raju, G., Wu, Z., and Weaver, P. (January, 2014)
On {Further} {Developments} of {Feasible} {Region} of {Lamination} {Parameters}
  for {Symmetric} {Composite} {Laminates}.
In \emph{55th {AIAA}/{ASME}/{ASCE}/{AHS}/{ASC} {Structures}, {Structural}
  {Dynamics}, and {Materials} {Conference}} National Harbor, Maryland: American
  Institute of Aeronautics and Astronautics.

\bibitem{2020Natella}
Natella, M.
Aeroelastic {Tailoring} of {Composite} {Aircraft}. (2020).

\bibitem{2021Rajpal}
Rajpal, D.
Dynamic aeroelastic optimization of composite wings including fatigue
  considerations. (2021).

\bibitem{2024Maathuis}
Maathuis, H.~F., De~Breuker, R., and Castro, S.~G. (January, 2024)
High-{Dimensional} {Bayesian} {Optimisation} with {Large}-{Scale} {Constraints}
  - {An} {Application} to {Aeroelastic} {Tailoring}.

\end{thebibliography}


\end{document}